\documentclass[a4paper,12pt]{article}
\usepackage[spanish,catalan,english]{babel}
\usepackage{amsfonts}

\newcommand{\zero}{\mathbf 0}
\newcommand{\Z}{\mathbb Z}

\newcommand{\F}{\mathbb F}

\newtheorem{theorem}{Theorem}

\newtheorem{coro}[theorem]{Corollary}

\newenvironment{demo}{\noindent {\bf Proof:}\ }{\ \rule{1mm}{2mm}}

\title{{Plotkin construction: Rank and Kernel}\thanks{
This work has been partially supported by the Spanish MEC and the
European FEDER MTM2006-03250 Grant and by the  PNL2006-13UAB
Grant.}}

\author{J. Borges, \ C. Fern\'{a}ndez \\ \\
     Departament d'Enginyeria de la Informaci\'{o} i de les Comunicacions \\
     Universitat Aut\`{o}noma de Barcelona \\
   08193 Bellaterra, Spain   }

\begin{document}

\maketitle

\begin{abstract}
Given two binary codes of length $n$, using Plotkin construction we
obtain a code of length $2n$. The construction works for linear and
nonlinear codes. For the linear case, it is straightforward to see
that the dimension of the final code is the sum of the dimensions of
the starting codes. For nonlinear codes, the rank and the dimension
of the kernel are standard measures of linearity. In this report, we
prove that both parameters are also the sum of the corresponding
ones of the starting codes.
\end{abstract}

\medskip


\section*{Introduction}
Let $\F^n$ be the vector space of dimension $n$ over $\Z_2$.  The
Hamming distance between vectors $x,y\in \F^n$, denoted by $d(x,y)$,
is the number of coordinates in which $x$ and $y$ differ. A binary
code $C$ of length $n$ is a subset of $\F^n$, its elements are
called codewords. The minimum distance of a code $C$ is the minimum
of the distances between pairs of different codewords.

Let $C$ be a binary code. If $C$ is a subspace of $\F^n$, then we
say that $C$ is a $[n,k,d]$ linear code, where $k$ is the dimension
of $C$ and $d$ is its minimum distance. If $C$ is not a subspace, we
say that is a nonlinear $(n,|C|,d)$ code.

Given a binary code $C$, the {\em rank} of $C$ is defined as the
dimension of the linear span of $C$. I.e. $rank(C)=dim \big< C
\big>$. The {\em kernel} of a binary code $C$, $Ker(C)$, is the set
of vectors that leave $C$ invariant under translation, i.e.
$Ker(C)=\{x\in\F^n \mid C+x=C\}$. If $C$ contains the all-zero
vector, then $Ker(C)$ is a linear subcode of $C$. Note that for a
linear code, the rank and the dimension of the kernel are simply the
dimension of the code.

Let $C_1$ and $C_2$ be two binary codes of length $n$. We can
construct a binary code $C$ of length $2n$ by terms of $C_1$ and
$C_2$ with the following construction:
\begin{equation}\label{eq:PlotkinC}
C=\{(u|u+v) : u\in C_1, v\in C_2 \},
\end{equation}
where `$|$' denotes concatenation. Such construction is called
Plotkin construction or $(u|u+v)$-construction and was first stated
by Plotkin in 1960 \cite{Plot}.

Let $C_1$ and $C_2$ be binary codes and let $C$ be the code obtained using
construction (\ref{eq:PlotkinC}).
\begin{itemize}
\item If $C_1$ and $C_2$ are linear $[n,k_1,d_1]$ and $[n,k_2,d_2]$ codes
then, $C$ is a linear $[2n,k_1+k_2,min\{2d_1,d_2\}]$ code.
\item If $C_1$ and $C_2$ are nonlinear $(n,|C_1|,d_1)$ and $(n,|C_2|,d_2)$ codes
then, $C$ is a nonlinear $(2n,|C_1|\cdot |C_2|,min\{2d_1,d_2\})$
code.
\end{itemize}

\begin{theorem}\label{theo:RK}
  Let $C_1$ and $C_2$ be binary codes of length $n$, $\zero \in C_1$,
  $\zero \in C_2$, and let $C$ be the binary code built from them via
  contruction (\ref{eq:PlotkinC}). Therefore,
\begin{itemize}
\item[(i)] $Ker(C)=\{(x|x+y) : x\in Ker(C_1), y\in Ker(C_2)\},$
\item[(ii)] $\big< C\big>=\{(x|x+y) : x\in \big< C_1\big>, y\in \big< C_2\big>\}.$
\end{itemize}
\end{theorem}

\begin{demo}
\begin{itemize}
\item[(i)]
For any $x\in Ker(C_1), y\in Ker(C_2), (u|u+v)\in C$ we obtain
$(x|x+y)+(u|u+v)=(x+u|(x+u)+(y+v))\in C$. Thus, $(x|x+y)\in Ker(C)$.

Now, let $(x|x+y)\in Ker(C)$. For any $u\in C_1$ and $v\in C_2$, we
have $(u|u+v)\in C$ and $(x|x+y)+(u|u+v)=(x+u|(x+u)+(y+v))\in C$. By
construction, $x+u\in C_1$ and, therefore, $x\in Ker(C_1)$. To prove
that $y\in Ker(C_2)$ we consider $v\in C_2$ and the codeword
$(x|x+v)\in C$. Then, since $(x|x+y) + (x|x+v)=(\zero|\zero +
(y+v))\in C$ we conclude that $y+v\in C_2$ and $y\in Ker(C_2)$.
\item[(ii)] Let $x=\sum_{i=0}^su_i\in \big< C_1\big>$, $y=\sum_{j=0}^tv_j \in
   \big< C_2\big>$, where $u_i\in C_1$, for $i=1,\dots,s$, $v_j\in C_2$, for
  $j=1,\dots,t$. Then
  $(x|x+y)=\sum_{i=0}^s(u_i|u_i+0)+\sum_{j=0}^t(0|0+v_j)\in
  \big<C\big>$.

Finally, if $(x|x+y)\in\big<C\big>$ then
$(x|x+y)=\sum_{i=0}^k(u_i|u_i+v_i)=(\sum_{i=0}^ku_i|(\sum_{i=0}^ku_i)+(\sum_{i=0}^kv_i))$,
where $x=\sum_{i=0}^ku_i\in \big< C_1\big>$ and $\sum_{i=0}^kv_i \in  \big< C_2\big>$.
\end{itemize}
\end{demo}

\begin{coro} With the same conditions as in Theorem \ref{theo:RK}
\begin{itemize}
\item[(i)] $dim(Ker(C))=dim(Ker(C_1))+dim(Ker(C_2)),$
\item[(ii)] $rank(C)=rank(C_1) + rank(C_2)$.
\end{itemize}
\end{coro}

\begin{demo}
It is easy to check that $f:Ker(C_1)\times Ker(C_2)\rightarrow
Ker(C)$, $f(x,y)=(x|x+y)$ and $g:\big< C_1\big> \times \big<
C_2\big>\rightarrow\big< C\big> $, $g(x,y)=(x|x+y)$ are bijections.
Therefore:
\begin{itemize}
\item[(i)] $|Ker(C)|=|Ker(C_1)|\cdot|Ker(C_2)|$ and
  $dim(Ker(C))=dim(Ker(C_1))+dim(Ker(C_2)).$
\item[(ii)] $|\big< C\big>|=|\big< C_1\big>|\cdot|\big< C_2\big>|$ and
   $rank(C)=rank(C_1) + rank(C_2)$.
\end{itemize}
\end{demo}



\begin{thebibliography}{zzzz99}



\bibitem{Plot} M. Plotkin. ``Binary codes with specified minimum
distances," {\em IEEE Trans. Inform. Theory}, vol. 6, pp. 445-450,
1960.

\end{thebibliography}
\end{document}